\documentstyle[11pt,AATS,epsf]{article}	

\def\gsimeq{\,\,\raise0.14em\hbox{$>$}\kern-0.76em\lower0.28em\hbox  
{$\sim$}\,\,}  
\def\lsimeq{\,\,\raise0.14em\hbox{$<$}\kern-0.76em\lower0.28em\hbox  
{$\sim$}\,\,} 

\def\gsimeq{\,\,\raise0.14em\hbox{$>$}\kern-0.76em\lower0.28em\hbox
{$\sim$}\,\,}
\def\lsimeq{\,\,\raise0.14em\hbox{$<$}\kern-0.76em\lower0.28em\hbox
{$\sim$}\,\,}
\def\chem#1#2{$\rm{}^{#1}\kern-0.8pt#2$}
\def\reac#1#2#3#4#5#6{$\rm\,{}^{#1}\kern-0.8pt{#2}\,({#3}\,,{#4})\,
{}^{#5}\kern-0.8pt{#6}\,$}

\begin{document}	

\title{Hope and Inquietudes in Nucleo-cosmochronology} 

\author{M. Arnould, S. Goriely}
\affil{Institut d'Astronomie et d'Astrophysique, Universit\'e Libre de Bruxelles,
B-1050 Brussels, Belgium}

\begin{abstract}
Critical views are presented on some nucleo-cosmo\-chronological questions.
Progress has been made recently in the development of the
\chem{187}{Re}-\chem{187}{Os} cosmochronometry. From this, there is good hope
for this clock to become of the highest quality for the nuclear dating of the
Universe. The {\it simultaneous} observation of Th and U in ultra-metal-poor
stars would also be a most interesting prospect. In contrast, a serious
inquietude is expressed about the reliability of the chronometric attempts
based on the classical \chem{232}{Th}-\chem{238}{U} and
\chem{235}{U}-\chem{238}{U} pairs, as well as on the Th (without U) abundance
determinations in ultra-metal poor stars.
\end{abstract}


\section{Introduction}

The use of radionuclides to estimate astrophysical ages has a long history,
a milestone of which is the much celebrated piece of work of Fowler
\& Hoyle (1960). For long, the field of nucleo-cosmochronology that has emerged
from this paper has been aiming at the determination of the age $T_{\rm nuc}$ of
the nuclides from abundances in the material making up the bulk of the solar
system. If indeed the composition of this material witnesses the long history of
the compositional evolution of the Galaxy prior to the isolation of the solar
material, a reliable evaluation of
$T_{\rm nuc}$ (clearly a lower bound to the age of the Universe) requires (i) the
identification of radionuclides with half-lives commensurable with estimated
reasonable galactic ages (i.e. $t_{1/2} \gsimeq 10^9$ y), (ii) the construction of
nucleosynthesis models that are  able to provide the isotopic or elemental yields
for these radionuclides, (iii) high quality data for the meteoritic abundances of
the  relevant nuclides, and, last but not least, (iv) the build-up of models
for the evolution of the abundances of these nuclides in the Galaxy,  primarily in
the solar neighborhood. All these requirements clearly make the chronometric task
especially demanding. While everybody would agree this far, there are different
ways to look at the question.

A pessimistic view is that nucleo-cosmochronology can at best set limits on
$T_{\rm nuc}$ through the use of a so-called `model-independent'
approach. Using this formalism, Meyer \& Schramm (1986) conclude that $9 \lsimeq
T_{\rm nuc} (\rm Gy) \lsimeq 27$ (only the lower limit is truly
model-independent. The upper limit depends on some model assumptions). This is
clearly not a highly constraining range!
An optimistic view is that, given the presumed complexity of the chemical
evolution of the galactic disk, it is by far preferable to describe its
nucleosynthetic history by a simple function with some adjustable parameters. This
has been advocated by Fowler over the years with the use of the so-called
`exponential model'.
A practitioner's view is that it is really worth studying nucleo-cosmochronology
in the framework of chemical evolution models which are simple enough not to
account for all the dynamical aspects of the formation of the present galactic
disk, but which imperatively satisfy as many observational constraints as possible
[e.g. Yokoi et al. (1983), Takahashi (1998)].

A new chapter in the story of nucleo-cosmochronology has been written with the
 discovery of isotopic anomalies attributed to the {\it in situ} decay in a minute
fraction of the meteoritic material of radionuclides with half-lives in the 
approximate $10^5 \lsimeq t_{1/2} \lsimeq 10^8$ y range. These observations
are hoped to provide some information on discrete nucleosynthesis  events
that  presumably contaminated the solar system at times between about $10^5$ and
$10^8$ y prior to the isolation of the solar material from the general  galactic
material. Constraints on the chronology of nebular and planetary events
in the early solar system could be gained concomitantly.  

Finally, the observation of Th in some very metal-poor stars and of U in one of
them has opened the way to a possible nuclear-based evaluation of the age of
individual stars other than the Sun. This clearly broadens the original scope of
nucleo-cosmochronology still further.

In Sect. 2, we reiterate the inquietude originally expressed by Yokoi et al.
(1983) concerning the classical \chem{232}{Th}-\chem{238}{U} and
\chem{235}{U}-\chem{238}{U} pairs, which may not be as reliable galactic clocks as
it is often stated. In contrast, we express some hope concerning
\chem{187}{Re}-\chem{187}{Os} in Sect. 3. Our second inquietude relates to the
chronological information one may really hope to gain from the observation of Th
in very low-metallicity stars (Sect. 4). The situation would be brighter if one
could rely on precise measurements of the Th/U ratio in such stars (Sect. 5). The
chronometry using short-lived radionuclides is not discussed here. The interested
reader is referred to e.g. Arnould et al. (2000) for a brief review of this
question and references (see also T. Lee, these proceedings).

\section{Inquietude 1: The trans-actinide clocks}
 
The most familiar long-lived \chem{232}{Th}-\chem{238}{U} and
\chem{235}{U}-\chem{238}{U} clocks based on the present meteoritic
content of these nuclides are reviewed by J.W. Truran (these Proceedings). At
several occasions, we have expressed reservations about the use of these pairs as
reliable evaluators of $T_{\rm nuc}$ [e.g. Arnould \& Takahashi (1990)]. 

The first reason for this inquietude relates to the clear necessity of
knowing with high precision the production ratios of the involved actinides. Such
quality predictions are clearly out of reach at the present time. One is indeed
dealing  with nuclides that can be produced by the r-process only, which suffers
from very many astrophysics and nuclear physics problems, as we have emphasized
at many  occasions [e.g. Arnould \& Takahashi 1999; Goriely \& Clerbaux (1999)].
In particular, the true astrophysical site(s) of the r-process remain(s) so far
unknown, preventing any firm (sometimes far-reaching) conclusions to be drawn in
relation with nucleosynthesis applications. On the nuclear physics side, the
nuclear properties (such as nuclear masses, deformations,
\dots) of thousands of exotic nuclei located between the  valley of 
$\beta$-stability and
the neutron drip line have to be known, as well as their transmutation rates
through $\alpha$- or $\beta$-decay, various fission channels, as well as
through nuclear reactions, such as ($n,\gamma$) and ($\gamma, n$). 
Despite much recent experimental effort, none of these quantities are known for the
nuclei involved in the r-process, so that they have to be extracted from theory. In
addition, the Th and U nuclides  are the only naturally-occurring ones
beyond \chem{209}{Bi}, so that any extrapolation relying  on semi-empirical analyses
and fits of the solar r-process abundance curve is in danger of being especially
unreliable. 

Recent r-process calculations (Goriely \& Clerbaux 1999) provide
production ratios in the
$0.8 \lsimeq P_{235}/P_{238} \lsimeq 1.2$ and $1.4 \lsimeq P_{232}/P_{238} \lsimeq
2.7$. The extent of these ranges largely forbids the build-up of precise
nucleo-cosmochronometries. It has to be noticed that this problem would linger even
if a realistic r-process model were given.
 
A second source of worry relates to the necessity of introducing
nucleo-cosmochronology in chemical evolution models of the Galaxy that satisfy at
best as many astronomical observables as possible, and to carefully check the
internal consistency of such extended chemical models. In fact, this step is a
quite complex one. Even if the nucleosynthesis yields of the radionuclides of
relevance are assumed to be reliably known for all stellar masses and
metallicities, which is far from being the case in reality, one still has to worry
about such effects as the so-called `astration', that is the possible more or less
substantial destruction during the stellar lifetime of those nuclei which were
absorbed from the interstellar medium at the stellar birth. From their constrained
one-zone model for the evolution of the composition of the galactic disk in the
solar neighborhood, Yokoi et al. (1983) conclude that (i)  the
predicted (\chem{235}{U}/\chem{238}{U})$_0$ and (\chem{232}{Th}/\chem{238}{U})$_0$
ratios at the time $T_\odot$ of isolation of the solar system material from the
galactic one about 4.6 Gy ago (here and in the following, the
subscript $0$ refers to this time) is only very weakly dependent on galactic ages,
at least in the explored range from about 11 to 15 Gy. This is analyzed as
resulting largely from the expected rather weak time dependence of the stellar
birthrate (except possibly at early galactic epochs, but a reliable information on
these times is largely erased by the subsequent long period of chemical
evolution), and (ii) the predicted abundance ratios at
$T_\odot$ are consistent with those derived at the same time from meteoritic
analyses if the r-process production ratios lie in the approximate ranges $1 <
P_{235}/P_{238} < 1.5$ and
$1.7 < P_{232}/P_{238} <2$. The latter values are consistent with the predicted
ones reported above, the former ones being only marginally so. At this point, it
has not to be forgotten that the estimate of the level of convergence between
predicted production ratios and observed abundances is blurred not only by the
large uncertainties in the estimated r-process actinide yields, but also by the
still quite large uncertainties affecting the meteoritic Th and U abundances, which
amount to at least 25\% and 8\%, respectively (Grevesse et al. 1996).

Finally, let us remind that most of the huge amount of work devoted to the
trans-actinide chronometries (e.g. J.W. Truran, these Proceedings) relies on the
simple exponential model whose substantial mathematical ease is obtained
at the expense of a quite scanty astrophysical content. This situation makes the
use of the Th-U chronologies especially unreliable. As examplified by Arnould \&
Takahashi (1990), a given exponential model may predict an increase by a factor of
3 in the $T_{\rm nuc}$ value by just changing the meteoritic
(\chem{232}{Th}/\chem{238}{U})$_0$ ratio from 2.32 to 2.67, which is compatible
with existing data (Grevesse et al. 1996)!  

From the above considerations, one may restate the conclusion already drawn by
Yokoi et al. (1983) that the relatively large uncertainties in the measured solar
(\chem{232}{Th}/\chem{238}{U})$_0$ and (\chem{235}{U}/\chem{238}{U})$_0$
ratios coupled with the relatively weak dependence with
galactic ages of these calculated ratios make it next to impossible to obtain a
{\it reliable} value of $T_{\rm nuc}$ from the trans-actinide chronometries as they
stand now. This conclusion holds even in the most favorable situation where the
r-process predictions can be made compatible with the abundance measurements in
the framework of a given model for the chemical evolution of the Galaxy. One has
to be aware of the fact that there is no guaranty at this time to reach this
necessary compatibility in a `natural' way, i.e. without having to play around
with a rich variety of parameters. 

\section{Hope 1: The \chem{187}{Re} - \chem{187}{Os} chronometry}
 
First introduced by Clayton (1964), the chronometry using the
\chem{187}{Re} - \chem{187}{Os} pair is able to avoid the difficulties 
related to the r-process modelling. True, \chem{187}{Re} is an r-nuclide.
However, \chem{187}{Os} is not produced  directly by the r-process, but 
indirectly via the $\beta^-$-decay of \chem{187}{Re} ($t_{1/2} 
\approx 42$ Gy) over the galactic lifetime. This makes it in principle
possible to derive a lower bound for $T_{\rm nuc}$
from the mother-daughter abundance ratio, provided that the `cosmogenic'
\chem{187}{Os} component is deduced from 
the solar abundance by subtracting its s-process contribution. This chronometry is
thus in the first instance reduced to a question concerning the s-process. 
This is a good news, as the s-process is, generally speaking, better under
nuclear or astrophysics control than the r-process. Other good news come from the
recent progress made in the measurement of the abundances of the
concerned nuclides in meteorites, which provide in addition the decay constant of
the neutral \chem{187}{Re} atoms. The derived value $\lambda = (1.666 \pm 0.010)
\time 10^{-11}$ y$^{-1}$ is substantially more precise than its direct
determination (Faestermann 1998). This improved input is essential for the
establishment of a reliable chronometry.  

Even if the s-process models are by far in much better shape than the r-process
ones, the evaluation of the relative production of \chem{186}{Os} and of 
\chem{187}{Os} by the s-process is not a trivial matter. One difficulty relates to
the fact that the \chem{187}{Os} 9.75 keV excited state can contribute
significantly to the stellar neutron-capture rate because of its thermal
population in s-process conditions ($T \gsimeq 10^8$ K). The ground-state capture
rate measured in the laboratory has thus to be modified. Several estimates of this
correction factor
$F_\sigma$ based on preliminary experimental data analyzed in the framework of
Hauser-Feshbach models are available  [e.g. McEllistrem et al. (1989)]. The
question has been re-examined with the code MOST (Goriely 1998). In particular, the
impact on the predictions of uncertainties in various key ingredients of the model
(like the $\Gamma_\gamma$-width or the neutron optical potential) has been analyzed
(Goriely, unpublished). There is also reasonable hope to reduce the uncertainty on
$F_\sigma$ through further dedicated experiments (Koehler and Mengoni, private
communications).

A second difficulty has to do with the possible branchings of the s-process path in
the $184 \leq A \leq 188$  region which may affect the evaluated
\chem{187}{Os}/\chem{186}{Os} s-process production ratio (Arnould et al. 1984).
The modelling of these branchings has been improved substantially with the
measurements of radiative neutron capture cross sections in the mass range of
relevance (K\"appeler et al. 1991), as well as thanks to detailed s-process
calculations in realistic model stars. We note in particular that estimate in the
framework of the proton-mixing scenario of thermally pulsing AGB stars (Goriely \& Mowlavi
2000) predict branching effects up to 20\% on the \chem{187}{Os}/\chem{186}{Os} ratio. 

The development of the Re - Os chronology also needs a reliable estimate of the
\chem{187}{Os} and \chem{187}{Re} yields from stars in a wide range of masses and
metallicities. This evaluation requires not only a good-quality modeling of the
s-process in order to predict the \chem{186}{Os}/\chem{187}{Os} ratio, but also of
other \chem{187}{Os} and \chem{187}{Re} transmutation channels in stellar interiors
which may affect the \chem{187}{Re}/\chem{187}{Os} ratio (`astration' effects). One
of these mechanisms is the possibility of enhanced
\chem{187}{Re} $\beta$-decay in stellar interiors. This is especially the result
of the bound-state $\beta$-decay phenomenon which has the dramatic effect of
reducing the \chem{187}{Re} half-life from about 42 Gy when one is dealing with a
neutral atom to a mere $32.9 \pm 2.0$ y when complete ionization is obtained. The
experiment that has allowed the measurement of this
half-life (Bosch 1996) is a significant step forward in the establishment of a
reliable Re - Os chronometry. It also put on safer grounds the evaluation of the
rate of transformation of \chem{187}{Os} into \chem{187}{Re} that can occur in
certain stellar layers through continuum electron captures (Arnould 1972,
Takahashi \& Yokoi 1983). A further complication arising in the evaluation of the
astration effects  comes from the fact that neutron captures can modify the
\chem{187}{Re}/\chem{187}{Os} ratio as well (Yokoi et al. 1983).

Clearly, the yield evaluation also requires the modelling of a variety of stars
with different masses and metallicities. Model stars adopted in recent studies of
the Re - Os chronology are briefly described by Takahashi (1998). The relevant yields
can also be found in his work.
 
Last but not least, it remains to construct a model for the chemical evolution of
the matter from which the solar system formed. This is the hardest part of all,
even in the framework of the simple constrained models already referred to
above. In their update of the study of Yokoi et al. (1983), Takahashi (1998) concludes
that the \chem{187}{Re} - \chem{187}{Os} pair leads to $T_{\rm nuc}$ values in the
reasonable $15 \pm 4$ Gy range. 

This result may not be as impressive as one may wish. However, there is good hope
that the Re - Os chronometry will turn well. Since the work of Yokoi et al.
(1983), much progress has been made on the nuclear physics and meteoritic sides,
and there is likely much more to come. This optimistic note cannot be expressed for
the trans-actinide clocks (Sect. 2). Of course, more sophisticated galactic
evolution models have to be devised. This is certainly of no small difficulty, but
this does not defy hope.
 
\section{Inquietude 2: Th in very metal-poor stars}

The recent observations of r-nuclides, including Th, in the metal-poor halo
stars CS 22892-052 or  HD115444 (see C. Sneden and J.W. Truran, these Proceedings)
have initiated a flurry of nucleo-cosmochronological excitement, both
observationally and theoretically. The search for r-process-rich very low
metallicity stars has been given an impressive boost. This is illustrated by the
analysis of CS 31082-001. This star is the first one whose U content (not just an
upper limit) has been measured (Cayrel et al. 2001). The abundances of other
r-nuclides have also been measured in this same star (V. Hill et al., these
Proceedings).  

On the theoretical side, it is now commonplace to claim that an ideal
nucleo-cosmochronometer has been found at last, the observed Th in the
above-mentioned stars providing a clean way of measuring the ages of these stars,
and consequently a reliable lower limit for the age of the Galaxy. In this
concerted optimism, some individuals (among whom the authors of this contribution)
dare expressing some reservation, however.

\subsection{Is the r-process pattern `universal'?}

{\it Almost} all astrophysicists dealing with the Th data in very low-metallicity
stars claim that the observed patterns of r-nuclide abundances are demonstrably
solar, implying a `universal' abundance distribution. This universality is of
essential importance for bringing the observed Th to the status of a chronometer.
It is clearly as essential not to take it too easily for granted!  

Goriely \& Arnould (1997) have tackled this question. They reach the conclusion
that the CS 22892-052 r-process composition in the best observationally
documented $56 \le Z \le 76$ range is remarkably close to the abundances obtained
from a distribution that fits very nicely the {\it whole} solar distribution.
However, they stress that it is very easy to construct `artificial' r-nuclide
distributions that reproduce almost equaly well the CS 22892-052 data in the $Z$
range mentioned above, while they diverge substantially from the solar data
outside of this range (including the actinides). Their so-called `random
distribution'  even appears to reproduce relatively well (and by far better than
the solar distribution) the observed $Z <56$ abundance distribution. 
  
From these results, Goriely \& Arnould (1997) [see
also Goriely \& Clerbaux (1999)] conclude that {\it the convergence of the solar,
CS 22892 - 052 and HD115444 abundance patterns in the $56 \le Z \le 76$ range does
in no way demonstrate the universality of the r-process, without excluding it,
however}. In fact, they stress that the r-pattern convergence is to a
large extent the signature of the nuclear properties of the r-progenitors of the
$56 \le Z \le 70$ elements, and does not provide any useful information on the
stellar conditions under which the r-process may develop.  This conclusion
fully applies as well to the recent abundance determinations of $50 \le Z \le 70$
elements in 22 metal-poor r-process-rich stars (Johnson \& Bolte 2001). In
fact, only observations of $A=90-130-195$ peak elements can shed light on
astrophysics issues concerning the r-process. 

In addition, there has been some bad news for the many proponents of the r-process
universality with a report on the preliminary analysis of the very metal-poor halo
star CS 31082 - 001, which shows large neutron-capture element
enhancements comparable to the ones observed in CS 22892 - 052 (V. Hill et al.,
these Proceedings). While a large similarity between the $56 < Z < 70$ element
patterns in CS 22892 - 052, HD115444 and CS 31082 - 001 is reported, the abundances in
the latter star differ significantly from those of the other two stars in the $Z > 70$
range. This includes Th, the Th/Eu ratio in CS 31082 - 001 being 2.8 times larger
than in CS 22892 - 052. Hence, under the universality assumption, CS 22892 - 052 (with
[Fe/H]=-3.1) is older than CS 31082 - 001 (with [Fe/H]=-2.9) by 21 Gy and would be
about 35 Gy old (see below). In addition of smashing to bits the idea of the
universality of the r-process which is so much beloved by some, the analysis of CS
31082 - 001 also reinforces our views that the question of the number of r-processes
(two being these days in favor in a substantial part of the nuclear physics
community) is of no substantial scientific interest. It would be more useful to
unravel the basic mysteries of the r-process before trying to establish the number of
r-process events!

\subsection{And if the r-process were indeed universal, would it help much the Th
chronometry?} 

From this point on, we will {\it assume} that the r-pattern of abundances is
universal. To make things clear, this means nothing more and nothing less than the
following: we assume that  {\it all possible blends of `r-process events' always
lead to the same final abundance pattern}, an event being defined here, as in the
canonical model of the r-process, by the ensemble ($T, N_n, t_{irr}$) of an
assumed constant temperature and neutron concentration during the irradiation time
$t_{irr}$. In direct relation with this statement, it may be worth making some
remarks: (i) calling for a blend of r-process events is not at all equivalent to
considering an ensemble of stars able to produce r-nuclides. As an example, a
single supernova is most likely the site of some continuum of r-process events;
(ii) the precise characteristics of the individual events which may contribute to
the universal mix are unknown, as well as the relative level of the contribution of
each of the events to the mix, and (iii) it cannot be excluded that different
combinations of different individual r-events lead to the same final
mix.\footnote{To make an analogy, it is well known in statistical mechanics that
each macroscopic state of a complex gas system can be obtained through a variety of
different superpositions of the states of the individual particles of the gas} 
 
Once the r-process universality is taken for granted, the procedure for estimating
the age of the Th-bearing metal-poor stars is straightforward, at least in its
principle. From the best possible fit to the solar system r-process abundance
distribution, the Th r-process production Th$_r$ can be deduced. The confrontation
between this value and the observed abundances leads trivially to the stellar ages.
Quite clearly, the consideration of very metal-poor stars allows to make the
economy of a model for the chemical evolution of the Galaxy. This very nice
feature is unfortunately compensated by the nightmare of evaluating Th$_r$ with the
high accuracy which is indispensable for building-up a chronometry.

As already mentioned in Sect. 2, the r-process remains
the most complicated nucleosynthetic process to model from the astrophysics as well as
nuclear physics point of view and is subject to large uncertainties that transpire
directly into the Th$_r$ predictions. The Th predictions are found to be especially
sensitive to the nuclear mass evaluations. As shown by Goriely \& Clerbaux (1999),
different mass models lead to stellar age differences that can amount to more than 20
Gy for {\it r-process models fitting equally well the r-process data for CS
22892 - 052 and HD 115444 in the $56 \leq Z \leq 82$ range}. The reader may wonder
why the quoted uncertainty is by far much larger than the one classically claimed
in the literature. The reason may be summarized as follows: it is a common
practice to identify the `best' nuclear mass models from a confrontation between
r-process predictions and the solar data (which are identical to the stellar ones
through the assumption of universality). This choice is biased, however. It
results indeed from implicit assumptions that are made concerning the
characteristic r-process events making up the universal mix. Other assumptions may
lead to different best nuclear models. In other words, it is meaningless to select
best mass models as long as the detailed characteristics of the suite of r-process
events contributing to an assumed universal r-process pattern remain unknown.

This is indeed clearly the case as even the proper site(s) for the r-process is
(are) not identified yet, all the proposed scenarios facing serious problems.
The astrophysics uncertainty that most critically affect the reliability of the
predictions of the Th production is clearly the still unknown maximum strength of
the neutron irradiation that can be obtained in an r-process event. Age
uncertainties amounting to about 16 Gy can result (Goriely \& Clerbaux 1999).

Finally, one does not have to underestimate the uncertainties that still affect
the evaluation of the contribution of the r-process to the solar-system Pb - Bi
peak, the predicted Th abundance being directly correlated to this contribution.
This situation is responsible for an uncertainty of about 20 Gy in the Th
chronometry (Goriely \& Clerbaux 1999).
 
As a conclusion, it is our opinion that the Th abundances observed in very
low-metallicity stars are of no chronometric virtues even if the universality of
the r-process is assumed, which is far from being demonstrated. As an example,
uncertainties in the age of CS 22892 - 052 just originating from the
errors in the Th abundance determination amounts to about 3.5 Gy. A
comparable precision of theoretical origin would impose a 15 - 20 \%  level of
accuracy in the r-process production of Th. This is just impossible to achieve!
In fact,  Goriely \& Clerbaux (1999) report an age for CS 22892-052 lying in the
$7 \lsimeq T^* {\rm[Gyr]} \lsimeq 39$ range! The many sources of uncertainties
briefly mentioned above indeed blur the picture substantially, at least if no undue
resort is made to too many `toothfairies'. 

\section{Hope 2: Th/U in very metal-poor stars} 

As already stated elsewhere (Arnould \& Takahashi 1999, Goriely \& Clerbaux
1999), a way to rescue the Th chronometry discussed in Sect. 4 would be to have
accurate measurements of the Th/U ratio in individual very low-metallicity stars.  
These nuclides are indeed likely to be produced concomitantly, so that one may
hope to be able to predict their production ratios more accurately than the
ratio of Th to any other r-element, in particular Eu.

A Th/U ratio has been reported recently for the star CS 31082 - 001 (Cayrel et al.
2001). This is real good news, even if the situation is not free of observational
and theoretical difficulties. The former ones are discussed by Cayrel et al.
(2001). On the theoretical side, one is still facing the severe
question of the universality of the r-process (Sect. 4.1.). In addition, even if
reduced, uncertainties remain in the evaluation of the Th/U production ratio.
Goriely \& Clerbaux (1999) estimate that it is likely to lie in the $1 \lsimeq
{\rm (Th/U)}_r \lsimeq 1.3$ range. This result, combined with the observed value
log$\epsilon {\rm(Th/U)}=0.74 \pm 0.15$ leads to an age for the considered star of
$14 \pm 3 \pm 2$ Gyr (where the errors correspond to observational and
theoretical uncertainties, respectively).

Clearly, the Th/U chronometry based on data for individual very metal-poor stars
has substantial potentialities which remain to be exploited. The detection and
analysis of other stars similar to CS 31082 - 001 would represent a major step
forward in this direction.

\section{Conclusion}

Nucleo-cosmochronology has the virtue of always being able to provide numbers one
can interpret as ages. The real challenge is to evaluate the reliability of these
predictions. We estimate that the \chem{187}{Re} - \chem{187}{Os} will turn well as a
chronometer based on meteoritic data. We are much more pessimistic concerning the
classical \chem{232}{Th} - \chem{238}{U} and \chem{235}{U} - \chem{238}{U} pairs. On
the other hand, we consider that it will remain close to impossible for long to date
individual stars in a secure way without the help of precise determinations of the
Th/U ratio in these stars. 


\acknowledgments S.G. is FNRS Research Associate.

\end{document}